\documentclass[journal,comsoc]{IEEEtran}
\usepackage{amsmath}
\usepackage{amssymb}
\usepackage{amsfonts}
\usepackage{graphicx}
\usepackage{epsfig}
\usepackage{subfigure}
\usepackage{psfrag}
\usepackage{xcolor}
\usepackage{epstopdf}
\usepackage{cite}
\usepackage[linesnumbered,ruled]{algorithm2e}

\begin{document}

\title{Wireless Power Meets Energy Harvesting: \\A Joint Energy Allocation Approach in OFDM-based System}

\author{Xun Zhou, Chin Keong Ho,~\IEEEmembership{Member,~IEEE}, and Rui Zhang,~\IEEEmembership{Member,~IEEE}
\thanks{This paper was presented in part at IEEE Global Conference on Signal and Information Processing (GlobalSIP), December, 2014, Atlanta, USA.}\thanks{X. Zhou is with the Department of Electrical and Computer Engineering, National University of Singapore (e-mail:xunzhou@u.nus.edu).}\thanks{C. K. Ho is with the Institute for Infocomm Research, A*STAR, Singapore (e-mail:hock@i2r.a-star.edu.sg).}\thanks{R. Zhang is with the Department of Electrical and Computer Engineering, National University of Singapore (e-mail:elezhang@nus.edu.sg). He is also with the Institute for Infocomm Research, A*STAR, Singapore.}}

\maketitle 

\begin{abstract}
This paper investigates an orthogonal frequency division multiplexing (OFDM)-based wireless powered communication system, where one user harvests energy from an energy access point (EAP) to power its information transmission to a data access point (DAP). The channels from the EAP to the user, i.e., the wireless energy transfer (WET) link, and from the user to the DAP, i.e., the wireless information transfer (WIT) link, vary over both time slots and sub-channels (SCs) in general. To avoid interference at DAP, WET and WIT are scheduled over orthogonal SCs at any slot. Our objective is to maximize the achievable rate at the DAP by jointly optimizing the SC allocation over time and the power allocation over time and SCs for both WET and WIT links. Assuming availability of full channel state information (CSI), the structural results for the optimal SC/power allocation are obtained and an offline algorithm is proposed to solve the problem. Furthermore, we propose a low-complexity online algorithm when causal CSI is available. 
\end{abstract}

\begin{IEEEkeywords}
Wireless power transfer, energy harvesting, wireless powered communication network (WPCN), orthogonal frequency division multiplexing (OFDM). 
\end{IEEEkeywords}

\IEEEpeerreviewmaketitle

\setlength{\baselineskip}{1.05\baselineskip}	
\newtheorem{definition}{\underline{Definition}}[section]
\newtheorem{fact}{Fact}
\newtheorem{assumption}{Assumption}
\newtheorem{theorem}{\underline{Theorem}}[section]
\newtheorem{lemma}{\underline{Lemma}}[section]
\newtheorem{corollary}{\underline{Corollary}}[section]
\newtheorem{proposition}{\underline{Proposition}}[section]
\newtheorem{example}{\underline{Example}}[section]
\newtheorem{remark}{\underline{Remark}}[section]
\newtheorem{conjecture}{\underline{Conjecture}}[section]
\newcommand{\mv}[1]{\mbox{\boldmath{$ #1 $}}}
\newcommand{\N}{\Gamma\sigma^2}
\newcommand{\Bmax}{B_{\rm max}}
\newcommand{\Dmax}{D_{\rm max}}
\newcommand{\RWF}{R_{\rm WF}}
\newcommand{\Sete}{\mathcal{N}_k^{\rm E}}
\newcommand{\Seti}{\mathcal{N}_k^{\rm I}}

\section{Introduction} 
In conventional wireless networks, nodes are powered by fixed energy sources, e.g., batteries. Finite network lifetime due to battery depletion becomes a fundamental bottleneck that limits the performance of energy-constrained networks, e.g., wireless sensor networks.
To prolong the operation time of the networks, the batteries have to be replaced or replenished manually after depletion. However, in some applications wireless nodes are deployed in conditions that replacement of batteries is inconvenient (e.g., for numerous sensors in large-scale sensor networks) or even infeasible (e.g., for implanted devices in human body). Alternatively, harvesting energy from renewable energy sources to power wireless devices becomes appealing by providing perpetual energy without battery replacement.

Wireless communications with energy harvesting (EH) transmitter in fading channels have been considered in \cite{Ho,Ozel}, where the throughput is maximized by energy allocation over time.
In contrast to a conventional transmitter with fixed power source, where the data transmission is adapted to the communication channels, the EH transmitter adapts its transmission both to the communication channels and to the dynamics of energy arrivals.
It is shown in \cite{Ho,Ozel} that when the battery at a transmitter has infinite storage, the optimal transmission power over slots follows a staircase water-filling (SWF) structure, where the water-levels (WLs) are nondecreasing over slots.
This is in contrast with the case of total energy constraint at the transmitter, in which the optimal transmission power is given by conventional water-filling (WF), where all slots share the same WL. The works \cite{Ho,Ozel} are extended to a two-hop relay network in \cite{CHuang,Gunduz,Gurakan,Orhan}, where both source and relay nodes employ EH to power their information transmission.

The energy sources in \cite{Ho,Ozel,CHuang,Gunduz,Gurakan,Orhan} are not dedicated or controllable, and are typically provided by the environment, such as solar energy, wind energy, thermal energy, and piezoelectric energy. Thus, the amount of harvested energy greatly depends on the conditions of the environment. With advances of radio frequency (RF) technologies, RF signals radiated from an access point (AP), referred to as {\it wireless power}, becomes a new viable source for EH. 
The radiation of RF signals is controllable, hence can potentially supply continuous and stable energy for wireless nodes. Utilizing wireless power as EH source to supply wireless nodes inspires research on {\it wireless energy transfer} (WET), the objective of which is to maximize the harvested energy at wireless nodes (see, e.g., \cite{Xu,Yang,Zeng,Zeng15June}). 

\begin{figure}
\centering
\includegraphics[width=8.5cm]{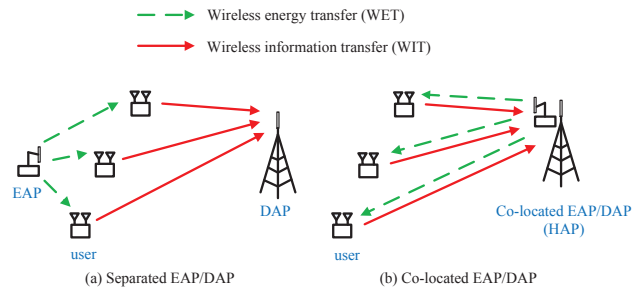} 
\caption{Architecture of wireless powered communication network (WPCN): separated EAP/DAP and co-located EAP/DAP.}
\label{fig:WPCN model} 
\end{figure}

Efficient WET by RF signals opens up the potential application of a new type of network, termed {\it wireless powered communication network} (WPCN) \cite{Bi}, where intended communication users are powered by dedicated wireless power. Fig. \ref{fig:WPCN model} illustrates the architecture of WPCN, where wireless users transmit information to a data access points (DAP) using the energy harvested from an energy access point (EAP). Hence, WET is performed at downlink from EAP to users, while wireless information transfer (WIT) is performed at uplink from users to DAP. For ease of practical implementation, each user is equipped with two antennas for EH and information transmission, respectively. In general, the DAP and EAP can be separately located in the network, referred to as the {\it separated EAP/DAP} case (see Fig. \ref{fig:WPCN model}(a)). A pair of DAP and EAP also can be co-located as a hybrid AP (HAP), providing the dual function of energy transfer and data access, which is referred to as the {\it co-located EAP/DAP} case (see Fig. \ref{fig:WPCN model}(b)). In both cases, the channel state information (CSI) of WET/WIT links is estimated at users and DAP, respectively, which are then sent to a central controller (located at EAP or DAP for example) for coordination of energy/information transmission. Separated EAP/DAP enjoys more flexibility in deployment of EAP/DAP; however, additional coordination and synchronization between the EAP and DAP is necessary. Co-located EAP/DAP is advantageous in information sharing (e.g., the channel estimation is simplified when uplink/downlink channel reciprocity applies) and hardware reuse (e.g., computational units). However, due to the same operation distance of WET and WIT for the co-located EAP/DAP case, the users far away from the HAP achieve low throughput, since higher transmission power needs to be consumed at these users yet with lower harvested energy, which is observed as a ``doubly near-far'' phenomenon in \cite{Ju}. 

For the case of co-located EAP/DAP, a harvest-then-transmit protocol is proposed in \cite{Ju} for a WPCN, where a HAP provides downlink energy transmission and uplink data access orthogonally in time to multiple EH users by the time division multiple access (TDMA) scheme. The work \cite{Ju} is extended to a full-duplex HAP setup in \cite{Ju-b}, where the downlink energy transmission and uplink information receiving is performed simultaneously at the HAP. 
In \cite{Sun}, the authors extend the work in \cite{Ju} by considering separated EAP/DAP, where the EAP is equipped with multiple antennas. In contrast to \cite{Ju}, which considers TDMA for uplink data transmission, space division multiple access (SDMA) is considered in \cite{Liu,Gang} by employing a multi-antenna HAP. Unlike \cite{Ju,Ju-b,Liu} which assume perfect CSI available at transmitters, \cite{Gang} studies the case of imperfect CSI by considering channel estimation.
Furthermore, \cite{Morsi} investigates the limiting distribution of the stored energy, the average error rate, and outage probability at the user when on-off transmission policy is adopted at the user assuming no CSIs for both WET/WIT links.
The capacity of large-size WPCN with geographically distributed users is studied in \cite{Huang} and \cite{Lee} for the separated EAP/DAP case, and in \cite{Che} for the co-located EAP/DAP case, based on the tool of stochastic geometry. Cooperative communication for WPCN is studied in \cite{Ju-c}, where near users help relay the information from far users to the HAP to overcome the doubly near-far problem.

Another line of work on joint wireless energy and information transmission has focused on the so-called SWIPT (simultaneous wireless information and power transfer) in downlink, which aims to characterize the achievable performance trade-off between harvesting energy and decoding information from the same signal waveform \cite{Zhang,Xun}. SWIPT is studied under various setups, e.g., in \cite{Zhang,Xu-b} for multiple-input multiple-output (MIMO) broadcast channels, in \cite{Xun-OFDM,Huang-OFDM,Derrick} for orthogonal frequency division multiplexing (OFDM) channels, and in \cite{Ding} for relay channels.

In this paper, we study an OFDM-based WPCN. Specifically, one user harvests energy from an EAP to power its information transmission to a DAP, while the EAP and DAP can be either separated or co-located (see Fig. \ref{fig:system model} for the separated case).
The total bandwidth of the system is equally divided into multiple orthogonal sub-channels (SCs). We consider energy/information transmission over finite time slots, where the channels of the WET and WIT links in general vary over both slots and SCs. 
Unlike \cite{Ju} where energy and information transmissions are performed over different time, in this paper we consider they are scheduled over orthogonal SCs at any slot. 
As a result, the user can harvest energy and transmit information at the same time, provided that the energy causality constraint \cite{Ho,Ozel} is satisfied, i.e., the total energy consumed for information transmission until any given time should be no greater than the total harvested energy so far. Moreover, the SCs are allocated separately for WET and WIT. The frequency orthogonality for WET/WIT inevitably introduces a tradeoff between energy and information transmissions. This additional design freedom via the SC allocation will influence the power allocation for the WET/WIT links.
Our objective is to maximize the achievable rate at the DAP by jointly optimizing the SC allocation over time and the power allocation over time and SCs for both WET and WIT links. The problem is investigated under two types of CSI availability for the WET and WIT links. The CSI can be either full CSI, which contains CSI of the past, present, and future slots, or causal CSI, which contains CSI only of the past and present slots. 

The main results of this paper are summarized as follows:
\begin{itemize}
\item Given full CSI, we prove that the optimal solution satisfies the following structural properties. First, at any given slot WET occurs at most on one SC. Second, given SC allocation WET can only occur in any {\it causally dominating} slot, i.e., a slot that has a larger channel power gain on the allocated SC than any of its previous slots. Third, when the initial battery energy of the user is zero, the optimal power allocation at the WIT link performs SWF over slots. That is, at any given slot the power allocated to different SCs has the same WL, while the WL increases after any causally dominating slot.
\item Given full CSI, we solve the problem by two stages. First, with given SC allocation, we obtain the optimal joint power allocation over time and SCs for both WET and WIT links. Second, we propose two heuristic schemes for the SC allocation obtained by a dynamic and a static SC allocation. 
\item Motivated by the optimal structural properties for the full CSI case, we propose an online scheme for the causal CSI case, namely, the scheme of dynamic SC with observe-then-transmit, which has low complexity.
\item For the full CSI case, our numerical results show that the performance by the proposed dynamic SC scheme is very close to the performance upper bound assuming non-interfering simultaneous WIT and WET over any SC. Furthermore, the numerical results demonstrate the superiority of WPCN with dedicated EAP over conventional EH system with random energy arrivals. For the causal CSI case, it is shown by numerical results that even utilizing partial information of channels for the WET link brings significant benefits to the achievable rate.
\end{itemize}

The rest of the paper is organized as follows. Section II introduces the system model. Section III presents the problem formulation. Section IV considers offline algorithm given full CSI. Section V proposes online algorithm given causal CSI. Section VI provides numerical examples. Finally, Section VII concludes the paper.

\section{System Model}\label{sec:system model} 
\begin{figure}
\centering
\includegraphics[width=8.5cm]{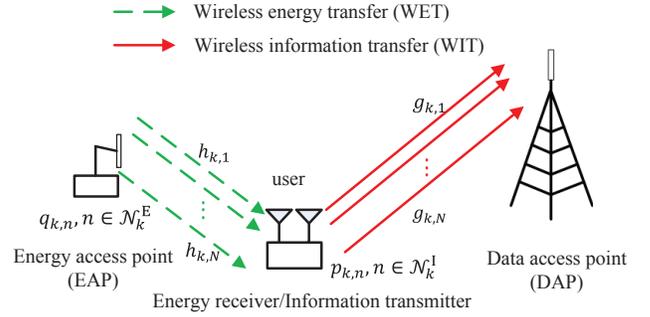} 
\caption{A wireless powered communication system where one user harvests energy from the EAP to power its information transmission to the DAP. The user is equipped with two antennas for energy harvesting and information transmission, respectively, over orthogonal sub-channels (SCs).}
\label{fig:system model} 
\end{figure}

We consider an OFDM-based wireless powered communication system, where one user harvests energy from an EAP to power its information transmission to a DAP (see Fig. \ref{fig:system model} for separated EAP/DAP). The EAP and DAP are each equipped with one antenna, while the user is equipped with two antennas. The EAP and DAP are connected to stable power supplies, whereas the user has no embedded energy sources.
Consider energy/information transmission in one block, which is equally divided into $K$ time slots, with each slot being of duration $T$. Let $T=1$ for convenience. The slots are indexed in increasing order by $k\in\mathcal{K}\triangleq\{1,\ldots,K\}$. 
The total bandwidth of the system is equally divided into $N$ orthogonal sub-channels (SCs). The SC set is denoted by $\mathcal{N}=\{1,\ldots,N\}$. The channel power gain from the EAP to the user, i.e., the WET link, during slot $k$ at SC $n$ is denoted by $h_{k,n}>0,k\in\mathcal{K},n\in\mathcal{N}$. The channel power gain from the user to the DAP, i.e., the WIT link, during slot $k$ at SC $n$ is denoted by $g_{k,n}>0,k\in\mathcal{K},n\in\mathcal{N}$. It is assumed that $h_{k,n}$'s and $g_{k,n}$'s are constant within one slot and SC, but vary over slots and SCs. In practice, for the co-located EAP/DAP case, $h_{k,n}$ and $g_{k,n}$ are correlated; while for the separated EAP/DAP case, $h_{k,n}$ and $g_{k,n}$ are independent. Our model is applicable for both scenarios.

Before the energy/information transmission during each slot $k$, the CSI of the WET and WIT links, i.e., $h_{k,n},g_{k,n},n\in\mathcal{N}$, is estimated. It is assumed that the channel estimation is sufficiently accurate such that the performance degradation due to the estimation error is negligible. Based on the CSI, energy/information transmission for the WET/WIT links is jointly scheduled.
To avoid interference at the DAP from the transmission signals by the EAP, WET and WIT are scheduled over orthogonal SCs\footnote{In practice, strict orthogonality of SCs imposes high requirements for hardware design. Energy leakage from one SC to adjacent SC may result in performance degradation, which is severe especially for the co-located EAP/DAP case as transmission power for the WET link is in general much larger than received power for the WIT link. For more detailed discussions, please refer to \cite{Naderi}.}. 
For notational simplicity, we define a dummy SC $n=0$, where $h_{k,0}=g_{k,0}=0$ for $k\in\mathcal{K}$, since there may be no SCs allocated for WET in slot $k$. The extended SC set is denoted by $\mathcal{N}'=\{0\}\cup\mathcal{N}$.
For each slot $k\in\mathcal{K}$, the SC set $\mathcal{N}'$ is partitioned into two complementary disjoint subsets for WET and WIT, denoted by $\Sete$ and $\Seti$, respectively, where $\Sete\subseteq\mathcal{N}',\Seti\subseteq\mathcal{N}'$, and $\Seti=\mathcal{N}'\backslash\Sete$.
The transmission power by the EAP during slot $k$ on SC $n$ is denoted by $q_{k,n}\geq0,k\in\mathcal{K},n\in\Sete$. The average transmission power at the EAP\footnote{In practice, the transmission power at each slot and SC may also be constrained by a peak power. In this paper, we assume the peak power constraint is sufficiently large compared to the transmission power, as WPCN in general operates at low power.} over $K$ slots in each block is denoted by $Q$, i.e.,  
\begin{equation}
\frac{1}{K}\sum_{k=1}^{K}\sum_{n\in\Sete} q_{k,n} \leq Q.
\end{equation}

The user harvests energy from the EAP by an energy receiver, and the energy is then stored in an energy buffer to power the information transmitter.
Assume the stored energy in the energy buffer at time instant $k^-$, i.e., the time instant just before slot $k$, is denoted by $B_k$. The initial energy at the buffer, i.e., $B_1$, is known. The transmission power by the information transmitter during slot $k$ at SC $n$ is denoted by $p_{k,n}\geq0,k\in\mathcal{K},n\in\Seti$. Assume the harvested energy during slot $k$ is ready for transmission at the end of slot $k$, the transmission power constraint at the user is thus given by
\begin{equation}\label{eq:transmit power constr org}
\sum_{n\in\Seti}p_{k,n}\leq B_k,k\in\mathcal{K}.
\end{equation}
We assume the storage of the energy buffer is sufficiently large compared to the harvested energy from the EAP, hence, no energy overflow at the energy buffer. We further assume except data transmission, other circuits at the user consume negligible energy. This is justified when data transmission consumes much larger power than that by other circuits, which is reasonable for most low-power (e.g., sensor) networks. The stored energy in the energy buffer at time instant $(k+1)^-$ is thus given by\footnote{Besides the radio signal from the EAP, the signal radiated by the information transmitter at the user can be a potential energy harvesting source for the energy receiver \cite{Zeng15}. However, the amount of the additional energy harvested from the information transmission is insignificant during the transmission block considered in this paper. This is because the power for information transmission is obtained from the harvested energy from the EAP, and the amount of additional energy harvested that can be recycled from the information transmission is insignificant, due to the pathloss between the two antennas at the user. For example, the pathloss can be −15dB, assuming an operation frequency centered at 900MHz with the distance of the two antennas as $\lambda/2$, where $\lambda$ denotes the wavelength of the transmission signal. In this case, the additional energy harvested from the information transmission is about 3$\%$ of the harvested energy from the EAP, which is negligible.
Hence, in this paper, we assume the energy harvested from the RF signals radiated by the information transmitter at the user is negligible as compared to energy harvested from the EAP. In practice, this amount of energy accumulated in a long term still may be saved for transmission in future blocks. In this case, the accumulated amount of energy can be utilized as the initial battery energy for the next transmission block.}
\begin{equation}\label{eq:battery user}
B_{k+1} = B_k+  \zeta\sum\limits_{n\in\Sete}h_{k,n}q_{k,n}-\sum\limits_{n\in\Seti}p_{k,n}, ~~ k\in\mathcal{K}
\end{equation}
where $\zeta$ denotes the energy efficiency at the user\footnote{In practice, the energy efficiency may be a nonlinear function of the received power by the antenna, depending on specific EH circuits implementation. In this paper, to simplify the system design, we assume the energy efficiency is constant and independent of the received power by the antenna.}, accounting for both conversion and discharging losses. Combine \eqref{eq:transmit power constr org} and \eqref{eq:battery user}, leading to the energy causality constraint
\begin{equation}
\sum\limits_{k=1}^i\sum\limits_{n\in\Seti} p_{k,n} \leq \zeta\sum\limits_{k=1}^{i-1}\sum\limits_{n\in\Sete} h_{k,n}q_{k,n}+B_1, ~ i\in\mathcal{K}.
\end{equation}

At the DAP, the receiver noise is modeled as a circularly symmetric complex Gaussian (CSCG) random variable with zero mean and variance $\sigma^2$. Due to frequency orthogonal transmission of energy and information, the energy signal from EAP will not interfere with the information reception at DAP. Moreover, the gap for the achievable rate from the channel capacity due to a practical modulation and coding scheme is denoted by $\Gamma\geq1$. The average achievable rate at the DAP in bps/Hz is thus
\begin{equation}
R=\frac{1}{KN}\sum\limits_{k=1}^K\sum\limits_{n\in\Seti} \log_2\left(1+\frac{g_{k,n}p_{k,n}}{\N}\right).
\end{equation}

\section{Problem Formulation}
Our objective is to maximize the average rate at the DAP by jointly optimizing the SC allocation over time and the power allocation over SCs and time for both WET and WIT links. This leads to the following optimization problem.
\begin{subequations}\label{prob:full CSI}
\begin{align}
&\mathop{\mathtt{max.}}_{\{\Sete\},\{q_{k,n}\},\{p_{k,n}\}}  ~ \frac{1}{KN}\sum\limits_{k=1}^K\sum\limits_{n\in\Seti} \log_2\left(1+\frac{g_{k,n}p_{k,n}}{\N}\right) \nonumber \\
&~~~~\mathtt{s.t.}  ~ \frac{1}{K}\sum\limits_{k=1}^{K}\sum\limits_{n\in\Sete} q_{k,n} \leq Q,  \tag{\theequation a}\label{eq:full CSI c1}\\
              & ~~~~~~~~ \sum\limits_{k=1}^i\sum\limits_{n\in\Seti} p_{k,n} \leq \zeta\sum\limits_{k=1}^{i-1}\sum\limits_{n\in\Sete} h_{k,n}q_{k,n}+B_1, i\in\mathcal{K}. \tag{\theequation b}\label{eq:full CSI c2}
\end{align}
\end{subequations}

The SCs for WIT are not explicit optimization variables because by definition a SC is used either for WIT or WET only. The case where a SC $n$ is neither used for WET nor WIT is covered by assigning it to be used for WET with $q_{k,n}=0$ or WIT with $p_{k,n}=0$. Since the energy harvested during the last slot $K$ is not available for any information transmission, without loss of optimality there should be no energy transmission at the last slot, i.e., $\mathcal{N}_K^{\rm E}=\{0\},\mathcal{N}_K^{\rm I}=\mathcal{N}$, as assumed henceforth.
The optimal solutions are denoted by $\{{\Sete}^\ast,k\in\mathcal{K}\}$, $\{q_{k,n}^\ast,k\in\mathcal{K},n\in{\Sete}^\ast\}$, $\{p_{k,n}^\ast,k\in\mathcal{K},n\in{\Seti}^\ast\}$, and the maximum average rate by $R^\ast$. 

Suppose that the SC allocation and power allocation for the WET are given, such that the constraint \eqref{eq:full CSI c1} is satisfied. Then Problem \eqref{prob:full CSI} is reduced to the conventional EH transmitter with energy arrivals $\left\{\sum_{n\in\Sete} h_{k,n}q_{k,n},k\in\mathcal{K}\right\}$ \cite{Ho,Ozel}. Hence, Problem \eqref{prob:full CSI} is more general with additional design freedoms available via the SC allocation and power allocation for the WET link, which will in turn influence the power allocation for the WIT link. 

We first solve Problem \eqref{prob:full CSI} assuming full CSI available in Section \ref{sec:full csi}. Based on the results for full CSI, we propose an online algorithm for Problem \eqref{prob:full CSI} under causal CSI in Section \ref{sec:causal csi}.

\section{Offline Algorithm for Joint SC and Power Allocation}\label{sec:full csi}
In this section, we consider Problem \eqref{prob:full CSI} when full CSI is available, where all the $h_{k,n}$'s and $g_{k,n}$'s are {\em a priori} known by a central controller at the beginning of each block transmission. Our aim is to study the structural properties of the optimal transmission policy, which will provide important insights. 
Given SC allocation, by Propositions \ref{proposition:A} and \ref{proposition:Aa}, we show that WET may occur only on the so-called causally dominating slots. Furthermore, Proposition \ref{proposition:Ab} shows that the power allocated for WET matches to the power consumed for WIT during the intervals between the causally dominating slots.
The insights will be used for developing heuristic online schemes when only casual CSI is available.

Given SC allocation $\Sete,k\in\mathcal{K}$, at slot $k$, the index of the best SC (i.e., the SC that has the largest channel power gain) for the WET link among SCs in $\Sete$, is denoted by $m(k)\in\mathcal{N}'$. Hence, 
\begin{equation}\label{eq:mk}
m(k)=\underset{\left\{n\in\mathcal{N}_k^{\rm E}\right\}}{\arg\max} ~ h_{k,n}.
\end{equation}
In the following proposition, we state that with given SC allocation, at each slot $k$ WET may only occur on the SC $m(k)$.
\begin{proposition}\label{proposition:A}
For Problem \eqref{prob:full CSI} with given SC allocation $\Sete,k\in\mathcal{K}$, we have $q_{k,n}^\ast=0$ for $n\neq m(k)$.
\end{proposition}
\begin{IEEEproof}
Please refer to Appendix \ref{appendix:proof proposition A}.
\end{IEEEproof}

The intuition of Proposition \ref{proposition:A} is as follows. Given SC allocation $\Sete,k\in\mathcal{K}$, consider energy allocation for the WET link at any slot $k$ with total energy $\sum_{n\in\Sete}q_{k,n}$. Since the harvested energy at the user increases linearly with $q_{k,n},n\in\Sete$, the harvested energy at the user is maximized by allocating all energy to $q_{k,m(k)}$, which has the largest $h_{k,n}$ for $n\in\Sete$.

By Proposition \ref{proposition:A}, at each slot it is optimal to allocate at most one SC from the set $\mathcal{N}'$ to perform WET, as the remaining SCs can be utilized for potential WIT. We define a SC allocation function $\Pi(k)\in\mathcal{N}'$ to denote the SC allocated for WET during slot $k,k\in\mathcal{K}$. Hence, $\Sete=\{\Pi(k)\},\Seti=\mathcal{N}'\backslash\{\Pi(k)\},k\in\mathcal{K}$. Note that $\Pi(k)$ can be assigned to the dummy SC $n=0$ in case there is no WET scheduled in slot $k$.
Problem \eqref{prob:full CSI} is then reformulated by the following problem.
\begin{subequations}\label{prob:full CSI re1}
\begin{align}
& \mathop{\mathtt{max.}}_{\{q_{k,\Pi(k)}\},\{p_{k,n}\},\{\Pi(k)\}} ~ \frac{1}{KN}\sum\limits_{k=1}^K\sum\limits_{n\in\Seti} \log_2\left(1+\frac{g_{k,n}p_{k,n}}{\N}\right) \nonumber \\
& ~~~~\mathtt{s.t.}  ~ \frac{1}{K}\sum\limits_{k=1}^{K}q_{k,\Pi(k)} \leq Q,  \tag{\theequation a}\label{eq:full CSI re1 c1}\\
			   & ~~~~~~~~ \sum\limits_{k=1}^i\sum\limits_{n\in\Seti} p_{k,n} \leq \zeta\sum\limits_{k=1}^{i-1}h_{k,\Pi(k)}q_{k,\Pi(k)}+B_1, i\in\mathcal{K}. \tag{\theequation b}\label{eq:full CSI re1 c2}
\end{align}
\end{subequations}

Problem \eqref{prob:full CSI re1} is non-convex due to the integer SC allocation function $\Pi(k),k\in\mathcal{K}$. Hence, we solve Problem \eqref{prob:full CSI re1} by two stages: we first solve Problem \eqref{prob:full CSI re1} with given SC allocation $\Pi(k),k\in\mathcal{K}$, where the joint power allocation for the WET/WIT links is optimized; next, we propose heuristic schemes for the SC allocation.

\subsection{Joint Power Allocation}

We first consider Problem (\ref{prob:full CSI re1}) with given $\Pi(k),k\in\mathcal{K}$, where we focus on the joint power allocation design for the WET/WIT links. Given SC allocation $\Pi(k),k\in\mathcal{K}$, then $\Sete,\Seti,k\in\mathcal{K}$ are known. For notational simplicity, when the SC allocation $\Pi(k),k\in\mathcal{K}$ is given in Problem \eqref{prob:full CSI re1}, we drop the subscript $\Pi(k)$ in $q_{k,\Pi(k)}$ and $h_{k,\Pi(k)}$, i.e., $q_k\triangleq q_{k,\Pi(k)}$, $h_k\triangleq h_{k,\Pi(k)}$ for $k\in\mathcal{K}$. We note that $h_k=0$ when $\Pi(k)=0,k\in\mathcal{K}$.

First, we investigate the properties for $\{q_k^\ast\}$ and $\{p_{k,n}^\ast\}$ for Problem \eqref{prob:full CSI re1} with given $\Pi(k),k\in\mathcal{K}$. 
To this end, given SC allocation $\Pi(k),k\in\mathcal{K}$, we define set $\mathcal{D}$ as follows
\begin{align}\label{eq:setD}
&\mathcal{D}  \triangleq  \{1,~{\rm if}~ \Pi(1)\in\mathcal{N}\} \nonumber\\
& ~ \cup\{k\in\{2,\ldots,K-1\}:\Pi(k)\in\mathcal{N},h_{k}> h_{j},\forall 1\leq j<k\}.
\end{align}
We note that for the slots in $\mathcal{D}$, the channel power gain $h_{k}$ is increasing with the slot index $k$; hence, the slots in $\mathcal{D}$ are called {\it causally dominating} slots.  For convenience, we index the elements in set $\mathcal{D}=\{d_1,d_2,\ldots,d_{|\mathcal{D}|}\}$ such that $d_i<d_j$ for $i<j$. The complementary set of $\mathcal{D}$ is denoted by $\mathcal{D}^{\rm c}$, i.e., $\mathcal{D}^{\rm c}=\mathcal{K}\backslash\mathcal{D}$.

We partition the slot set $\mathcal{K}$ for the WIT link into subsets $\mathcal{D}_i\triangleq\{d_{i-1}+1,\ldots,d_{i}\},i=1,\ldots,|\mathcal{D}|+1$, referred to as the $i$th {\it interval}, where we set $d_0=0$ and $d_{|\mathcal{D}|+1}=K$ for notational simplicity. Thus, $\bigcup_i\mathcal{D}_i=\mathcal{K}$ and $\mathcal{D}_i\cap\mathcal{D}_j=\emptyset$ for $i\neq j$. In the following proposition, we show that WET only occurs in the causally dominating slots in $\mathcal{D}$.
\begin{proposition}\label{proposition:Aa}
For Problem \eqref{prob:full CSI re1} with given $\Pi(k),k\in\mathcal{K}$, the optimal power allocation satisfies $q_k^\ast=0$ for $k\in\mathcal{D}^{\rm c}$.
\end{proposition}
\begin{IEEEproof}
Please refer to Appendix \ref{appendix:proof proposition Aa}.
\end{IEEEproof}

\begin{remark}\label{remark:sparse}
Proposition \ref{proposition:Aa} shows that WET occurs {\em sparsely} in time, i.e., WET occurs only when a slot dominating all its previous slots. Intuitively, this is because instead of allocating energy to any slot in $\mathcal{D}^{\rm c}$, allocating the same amount of energy to an earlier slot in $\mathcal{D}$ which has larger channel power gain at the WET link will result in a larger feasible region for $\{p_{k,n}\}$. 
\end{remark}

Further in Proposition \ref{proposition:Ab}, it is shown that if the energy at the user is used up after a particular casually dominating slot, then the energy is used up after later causally dominating slots. 
\begin{proposition}\label{proposition:Ab}
In Problem \eqref{prob:full CSI re1} with given $\Pi(k),k\in\mathcal{K}$, if $\{q_k^\ast\}$ and $\{p_{k,n}^\ast\}$ satisfy 
\begin{equation}\label{eq: first equal A}
\sum\limits_{k=1}^{d_j}\sum\limits_{n\in\Seti} p_{k,n}^\ast =\zeta\sum\limits_{k=1}^{d_j-1} h_{k}q_k^\ast+B_1
\end{equation}
where $d_1\leq d_j\leq d_{|\mathcal{D}|}$, i.e., constraint \eqref{eq:full CSI re1 c2} holds with equality at $i=d_j$, then we have
\begin{equation}
\sum\limits_{k\in\mathcal{D}_{l+1}}\sum\limits_{n\in\Seti} p_{k,n}^\ast = \zeta h_{d_{l}}q_{d_{l}}^\ast, ~~ l=j,\ldots,|\mathcal{D}|.
\end{equation}
\end{proposition}
\begin{IEEEproof}
Please refer to Appendix \ref{appendix:proof proposition Ab}.
\end{IEEEproof}

Next, we discuss two cases for the initial battery energy $B_1$, i.e., the special case of $B_1=0$ and the general case of $B_1\geq0$.
\subsubsection{Zero Initial Battery Energy with $B_1=0$}
We first consider the case $B_1=0$. With $B_1=0$, from Proposition \ref{proposition:Aa} and \eqref{eq:full CSI re1 c2}, we have
\begin{equation}\label{eq:pkn D1}
\sum_{k=1}^{d_1}\sum_{n\in\Seti} p_{k,n}=0.
\end{equation}  
Thus, 
\begin{equation}\label{eq:pkn 0}
p_{k,n}^\ast=0, ~~ k\in\mathcal{D}_1, n\in\Seti
\end{equation}
From Proposition \ref{proposition:Aa}, $q_k^\ast=0,k\in\mathcal{D}^{\rm c}$.
Henceforth, we consider optimization for $\{p_{k,n},k=d_1+1,\ldots,K,n\in\Seti\}$ and $\{q_k,k\in\mathcal{D}\}$.

From \eqref{eq:pkn D1}, constraint (\ref{eq:full CSI re1 c2}) holds with equality at $i=d_1$, from Proposition \ref{proposition:Ab} we have 
\begin{equation}
\sum\limits_{k\in\mathcal{D}_{l+1}}\sum\limits_{n\in\Seti} p_{k,n}^\ast = \zeta h_{d_{l}}q_{d_{l}}^\ast, ~~ l=1,\ldots,|\mathcal{D}|.
\end{equation}
Define the {\it effective channel power gain} as
\begin{align}\label{eq:gi'-define A}
g_{k,n}'&=h_{d_i}g_{k,n}, k\in\mathcal{D}_{i+1},i=1,\ldots,|\mathcal{D}|,n\in\Seti
\end{align}
Define $\{p_{k,n}'\}$ as
\begin{align}\label{eq:pi'-define A}
p_{k,n}'&=p_{k,n}/h_{d_i}, k\in\mathcal{D}_{i+1},i=1,\ldots,\mathcal{|D|},n\in\Seti.
\end{align}
From \eqref{eq:pkn 0}-\eqref{eq:pi'-define A}, Problem \eqref{prob:full CSI re1} with given $\Pi(k),k\in\mathcal{K}$ and $B_1=0$ is equivalent to the following problem.
\begin{subequations}\label{prob:sub1}
\begin{align}
\mathop{\mathtt{max.}}_{\{q_{d_i}\},\{p_{k,n}'\}} &  \frac{1}{KN}\sum\limits_{k=d_1+1}^{K}\sum\limits_{n\in\Seti}\log_2\left(1+\frac{g_{k,n}'p_{k,n}'}{\N}\right) \nonumber \\
\mathtt{s.t.} &  \sum\limits_{k=d_1+1}^{K}\sum\limits_{n\in\Seti} p_{k,n}'=\zeta KQ,  \tag{\theequation a}\\
              &  \sum\limits_{k\in\mathcal{D}_{i+1}}\sum\limits_{n\in\Seti} p_{k,n}'=\zeta q_{d_i}, i=1,\ldots,|\mathcal{D}|.\tag{\theequation b}\label{eq:sub1 c2}
\end{align}
\end{subequations}
We recognize that the optimization over $\{p_{k,n}',k=d_1+1,\ldots,K,n\in\Seti\}$ is then
a water-filling (WF) problem over time slots $k\in\{d_1+1,\ldots,K\}$ and SCs $n\in\Seti$, because $\{q_{d_i}\}$ can be arbitrarily chosen and thus the last constraint becomes redundant.
The optimal $\{p_{k,n}',k=d_1+1,\ldots,K,n\in\Seti\}$ is then obtained by the so-called WF power allocation over slots/SCs, given by
\begin{equation}\label{eq:pi'-optimal A}
  p_{k,n}'=\left(\frac{1}{\lambda KN\ln2}-\frac{\N}{g_{k,n}'}\right)^+,  k=d_1+1,\ldots,K
\end{equation}
where $(a)^+\triangleq\max(0,a)$, and $\lambda$ satisfies $\sum_{i=d_1+1}^{K}\sum_{n\in\Seti}p_{k,n}'=\zeta KQ$. The water-level (WL) is given by $(\lambda KN\ln2)^{-1}$.
From \eqref{eq:pkn 0}, (\ref{eq:pi'-define A}), and (\ref{eq:pi'-optimal A}), the optimal $\{p_{k,n}^\ast,k\in\mathcal{K},n\in\Seti\}$ is given by
\begin{align}\label{eq:pi-opt A}
p_{k,n}^\ast=\begin{cases}
0, & k\in\mathcal{D}_1,\\
\left(\frac{h_{d_i}}{\lambda KN\ln2}-\frac{\N}{g_{k,n}}\right)^+, & k\in\mathcal{D}_{i+1},  i=1,\ldots,|\mathcal{D}|.
\end{cases}
\end{align}
From \eqref{eq:sub1 c2} and Proposition \ref{proposition:Aa}, the optimal $\{q_k^\ast,k\in\mathcal{K}\}$ is given by
\begin{align}\label{eq:qi B1=0}
q_j=\begin{cases}
\frac{1}{\zeta}\sum\limits_{k\in\mathcal{D}_{i+1}}\sum\limits_{n\in\Seti}p_{k,n}', & j=d_i, i=1,\ldots,|\mathcal{D}|, \\
0, & {\rm otherwise}.
\end{cases}
\end{align}

\begin{remark}
From (\ref{eq:pi-opt A}), the optimal power allocation for the WIT link is adaptive to channels for both WET and WIT links. Moreover, the WLs are the same for slots and SCs in the same interval, while the WL for interval $\mathcal{D}_{i+1}$ is increasing over index $i$. Thus, the power allocation for the WIT link performs staircase water-filling (SWF) over slots.
\end{remark}

\subsubsection{Arbitrary Initial Battery Energy with $B_1\geq0$}
Now we consider the case with arbitrary initial battery energy at the user, i.e., $B_1\geq0$.

We note that in Problem \eqref{prob:full CSI re1} with given $\Pi(k),k\in\mathcal{K}$, $\{q_k^\ast\}$ and $\{p_{k,n}^\ast\}$ satisfy constraint \eqref{eq:full CSI re1 c2} with equality at the last slot $K=d_{|\mathcal{D}|+1}$; otherwise, the objective function can be increased by increasing some $p_{K,n}$. 
Let $d_x, 1\leq x\leq |\mathcal{D}|+1$, denote the first slot index in $\mathcal{D}\cup\{K\}$ such that $\{q_k^\ast\}$ and $\{p_{k,n}^\ast\}$ satisfy constraint (\ref{eq:full CSI re1 c2}) with equality. Hence, 
\begin{align}
& \sum\limits_{k=1}^{d_i}\sum\limits_{n\in\Seti}p_{k,n}^\ast < \zeta\sum\limits_{k=1}^{i-1}h_{d_k}q_{d_k}^\ast+B_1, i=1,\ldots,x-1, \label{eq:C2-less A}\\
& \sum\limits_{k=1}^{d_x}\sum\limits_{n\in\Seti}p_{k,n}^\ast = \zeta\sum\limits_{k=1}^{x-1}h_{d_k}q_{d_k}^\ast+B_1\label{eq:C2-dx A}
\end{align}
where we define $h_0\triangleq 1$ and $q_0\triangleq 0$.

\begin{lemma}\label{lemma:Aa}
For Problem \eqref{prob:full CSI re1} with given $\Pi(k),k\in\mathcal{K}$, $q_{d_k}^\ast=0$ for $k<x-1$. The optimal $\{q_k^\ast\}$ and $\{p_{k,n}^\ast\}$ satisfy
\begin{align}\label{eq:lemma Aa}
\sum\limits_{k=1}^{d_x}\sum\limits_{n\in\Seti}p_{k,n}^\ast=\zeta h_{d_{x-1}}q^\ast_{d_{x-1}}+B_1.
\end{align} 
\end{lemma}
\begin{IEEEproof}
Please refer to Appendix \ref{appendix:proof lemma Aa}.
\end{IEEEproof}

Similar to the case of $B_1=0$, we define the {\it effective channel power gain} as
\begin{align}\label{eq:gi'-caseI A}
g_{k,n}'=\begin{cases}
h_{d_{x-1}}g_{k,n}, & k = 1,\ldots,d_{x},n\in\Seti, \\
h_{d_i}g_{k,n}, & k\in\mathcal{D}_{i+1},i=x,\ldots,|\mathcal{D}|,n\in\Seti.
\end{cases}
\end{align} 
Define $\{p_{k,n}'\}$ as 
\begin{align}\label{eq:pi'-caseI A}
p_{k,n}'=\begin{cases}
p_{k,n}/h_{d_{x-1}}, & k = 1,\ldots,d_{x},n\in\Seti, \\
p_{k,n}/h_{d_i}, & k\in\mathcal{D}_{i+1},i=x,\ldots,|\mathcal{D}|,n\in\Seti.
\end{cases}
\end{align}

In the following lemma, we show that Problem \eqref{prob:full CSI re1} with given $\Pi(k),k\in\mathcal{K}$ is equivalent to a problem with optimizing variables $\{p_{k,n}'\}$.
\begin{lemma}\label{lemma:Ab}
Problem \eqref{prob:full CSI re1} with given $\Pi(k),k\in\mathcal{K}$ is equivalent to the following problem.
\begin{subequations}\label{prob:full CSI re2}
\begin{align}
\mathop{\mathtt{max.}}_{\{p_{k,n}'\}} & ~~ \frac{1}{KN}\sum\limits_{k=1}^{K}\sum\limits_{n\in\Seti}\log_2\left(1+\frac{g_{k,n}'p_{k,n}'}{\N}\right) \nonumber\\
\mathtt{s.t.} & ~~ \sum\limits_{k=1}^{K}\sum\limits_{n\in\Seti} p_{k,n}'\leq \zeta KQ+\frac{B_1}{h_{d_{x-1}}},  \tag{\theequation a}\label{eq:full CSI re2 c1}\\
			  & ~~ \sum\limits_{k=1}^{d_{x-1}}\sum\limits_{n\in\Seti}p_{k,n}' \leq \frac{B_1}{h_{d_{x-1}}},\tag{\theequation b}\label{eq:full CSI re2 c2}\\
              & ~~ \sum\limits_{k=1}^{d_x}\sum\limits_{n\in\Seti}p_{k,n}' \geq \frac{B_1}{h_{d_{x-1}}}. \tag{\theequation c}\label{eq:full CSI re2 c3}
\end{align}
\end{subequations}
The optimal $\{p_{k,n}^\ast\}$ is obtained by (\ref{eq:pi'-caseI A}); the optimal $\{q_k^\ast\}$ is obtained by
\begin{align}\label{eq:qi opt A}
q_j=\begin{cases}
\frac{1}{\zeta}\left(\sum\limits_{k=1}^{d_x}\sum\limits_{n\in\Seti}p_{k,n}'-\frac{B_1}{h_{d_{x-1}}}\right), & j=d_{x-1}, \\
\frac{1}{\zeta}\sum\limits_{k\in\mathcal{D}_{i+1}}\sum\limits_{n\in\Seti}p_{k,n}', & j=d_i, i=x,\ldots,|\mathcal{D}|, \\
0, & {\rm otherwise}.
\end{cases}
\end{align} 
\end{lemma}
\begin{IEEEproof}
Please refer to Appendix \ref{appendix:proof lemma Ab}.
\end{IEEEproof}
 
Problem (\ref{prob:full CSI re2}) is solved by the following proposition.
\begin{proposition}\label{proposition:Ac}
For Problem (\ref{prob:full CSI re2}), the optimal $\{p_{k,n}',k\in\mathcal{K},n\in\Seti\}$ is either given by
\begin{align}
p_{k,n}'=\left(\frac{1}{\lambda KN\ln 2}-\frac{\N}{g_{k,n}'}\right)^+
\end{align}
where $\lambda$ satisfies $\sum_{k=1}^K\sum_{n\in\Seti} p_{k,n}'=\zeta KQ+{B_1}/{h_{d_{x-1}}}$; or given by
\begin{align}\label{eq:pi' opt A}
p_{k,n}'=\begin{cases}
\left(\frac{1}{(\lambda-\mu)KN\ln 2}-\frac{\N}{g_{k,n}'}\right)^+, & k = 1,\ldots,d_{x}, \\
\left(\frac{1}{\lambda KN\ln 2}-\frac{\N}{g_{k,n}'}\right)^+, & k=d_x+1,\ldots,K.
\end{cases}
\end{align}
where $\lambda$ and $\mu$ satisfy $\sum_{k=1}^{d_x}\sum_{n\in\Seti}p_{k,n}'={B_1}/{h_{d_{x-1}}}$ and $\sum_{k=d_x+1}^{K}\sum_{n\in\Seti}p_{k,n}'=\zeta KQ$.
\end{proposition}
\begin{IEEEproof}
Please refer to Appendix \ref{appendix:proof proposition Ac}.
\end{IEEEproof}

To summarize, Problem \eqref{prob:full CSI re1} given $\Pi(k),k\in\mathcal{K}$ can be solved as follows: for each $d_x,1\leq x\leq |\mathcal{D}|+1$,
solve Problem (\ref{prob:full CSI re2}) to obtain $\{q_k\}$, $\{p_{k,n}\}$, and the objective value, denoted by $R(d_x)$. The optimal $d_x$ is then obtained by the $d_x$ which achieves the largest rate $R(d_x)$ and the corresponding power allocation $\{q_k\}$ and $\{p_{k,n}\}$ satisfy the constraints (\ref{eq:full CSI re1 c1}) and (\ref{eq:full CSI re1 c2}).
We propose Algorithm \ref{algorithm:1A} to solve Problem \eqref{prob:full CSI re1} with given $\Pi(k),k\in\mathcal{K}$.
\begin{algorithm}
\caption{Algorithm for solving Problem \eqref{prob:full CSI re1} with given $\Pi(k),k\in\mathcal{K}$.}
\label{algorithm:1A}
\KwIn{number of slots $K$; number of SCs $N$; initial battery energy $B_1$; channel power gain for the WET/WIT links $\{h_{k,n}\}$ and $\{g_{k,n}\}$; SC allocation $\Pi(k),k\in\mathcal{K}$}
\KwOut{optimal value $R(d_x^\ast)$; optimal power allocation $\{q_k^\ast\}$ and $\{p_{k,n}^\ast\}$}
\BlankLine
\For{each $x=1,\ldots,|\mathcal{D}|+1$}{
	Set effective channel power gain $\{g_{k,n}'\}$ by (\ref{eq:gi'-caseI A})\; 	
	Obtain $p_{k,n}',k\in\mathcal{K},n\in\Seti$ by WF algorithm with total power $\zeta KQ+{B_1}/{h_{d_{x-1}}}$\;
	\If{$\sum\limits_{k=1}^{d_x}\sum\limits_{n\in\Seti}p_{k,n}'<{B_1}/{h_{d_{x-1}}}$}{
		Obtain $p_{k,n}',k=1,\ldots,d_x,n\in\Seti$ by WF algorithm with total power ${B_1}/{h_{d_{x-1}}}$\; 
		Obtain $p_{k,n}',k=d_x+1,\ldots,K,n\in\Seti$ by WF algorithm with total power $\zeta KQ$\;
	}
	Obtain $\{p_{k,n}\}$ and $\{q_{k}\}$ by (\ref{eq:pi'-caseI A}) and (\ref{eq:qi opt A}), respectively, and obtain the corresponding rate $R(d_x)$\; \label{step A}
	\If{$\{q_k\}$ and $\{p_{k,n}\}$ do not satisfy (\ref{eq:full CSI re1 c1}) or (\ref{eq:full CSI re1 c2})}{
		Set $R(d_x)$ as zero\;
	}
}
Set $d_{x}^\ast=\underset{d_x}{\arg\max} ~R(d_{x})$\; 
The achievable rate for Problem \eqref{prob:full CSI re1} with given $\Pi(k),k\in\mathcal{K}$ is given by $R(d_x^\ast)$. The optimal power allocation $\{q_k^\ast\}$ and $\{p_{k,n}^\ast\}$ are obtained by step \ref{step A} correspond to the $x^\ast$th iteration.
\end{algorithm}

\subsection{SC Allocation}

Next, we consider the SC allocation design for Problem \eqref{prob:full CSI re1}, i.e., the integer function $\Pi(k),k\in\mathcal{K}$. The optimization on the integer function $\Pi(k),k\in\mathcal{K}$ is non-convex. In general, the complexity of exhaustive search over all possible $\Pi(k),k\in\mathcal{K}$ is $\mathcal{O}(N^K)$. 
Hence, we propose heuristic schemes for the SC allocation, which are easy to implement in practice, namely the dynamic SC scheme and the static SC scheme.

Define a SC allocation function $\tilde{\Pi}(k)$, which allocates the best SC for the WET link among all SCs $\mathcal{N}'$ at each slot $k$ for WET, i.e.,
\begin{equation}\label{eq:best SC}
\tilde{\Pi}(k)=\underset{n\in\mathcal{N}'}{\arg\max}~h_{k,n}, ~ k\in\mathcal{K}.
\end{equation}
Let $\tilde{\mathcal{D}}$ denote the causally dominating slot set obtained by \eqref{eq:setD} given SC allocation $\tilde{\Pi}(k)$. From Proposition \ref{proposition:Aa}, WET should occur only at causally dominating slots, hence, we let $\Pi(k)=0$ for $k\in\tilde{D}^{\rm c}$ such that potential information transmission can be performed at SCs $\tilde{\Pi}(k),k\in\tilde{D}^{\rm c}$.
In the {\em dynamic SC scheme}, the SC allocation is then given by
\begin{align}\label{eq:prior WET SC}
\Pi(k)=\begin{cases}
\tilde{\Pi}(k), & k\in\tilde{\mathcal{D}},\\
0, & {\rm otherwise}.
\end{cases}
\end{align}

\begin{remark}\label{remark:upper bound}
In Problem \eqref{prob:full CSI re1}, a performance upper bound for any SC allocation is obtained by allowing energy and information to transmit simultaneously using the same SC, while employing perfect interference cancellation at the DAP. Mathematically, this is equivalent to letting $\Sete=\Seti=\mathcal{N}',k\in\mathcal{K}$ in Problem \eqref{prob:full CSI}, which is then solved by the following lemma.
\end{remark}
\begin{lemma}\label{lemma:upper bound}
Problem \eqref{prob:full CSI} with $\Sete=\Seti=\mathcal{N}',k\in\mathcal{K}$ achieves same rate as Problem \eqref{prob:full CSI re1} with $\Pi(k)$ given in \eqref{eq:prior WET SC} and $\Seti=\mathcal{N}',k\in\mathcal{K}$.
\end{lemma}
\begin{IEEEproof}
Please refer to Appendix \ref{appendix:proof lemma ub}.
\end{IEEEproof}

In the {\em static SC scheme}, one SC is selected and fixed for WET throughout the whole transmission block, i.e., $\Pi(k)=n,k\in\mathcal{K}$, where the optimal choice of $n$ is obtained by exhaustive search over the SC set $\mathcal{N}'$ and selecting the one which achieves the largest rate. Therefore, the complexity of exhaustive search over all possible $\Pi(k)=n,k\in\mathcal{K}$ is $\mathcal{O}(N)$.

\section{Online Algorithm for SC Allocation}\label{sec:causal csi}

In this section, we consider online algorithms when causal CSI is available. In general, online algorithms can be designed optimally based on dynamic programming (DP) \cite{Ho}. However, the DP approach usually involves recursive computation with high computing complexity, which may be complicated for practical implementation. Furthermore, DP requires knowledge of channel statistics, e.g., the joint probability density function of the channel power gains for the WET/WIT links, which may be non-stationary or not available. Therefore, we aim to design online algorithm that has low complexity and requires only the past and present channel observations. 
Motivated by the results for the full CSI case, our online algorithm partitions the transmission block into subsets, and perform WET on the expected best SC in each subset. In particular, a simple scheme is proposed for the SC selection, which requires channel observations only of the past and present slots for the WET link.

For the full CSI case (assuming zero initial battery energy), the transmission block is partitioned as intervals according to the channels for the WET link, and the information transmission during each interval $\mathcal{D}_{i+1},i=1,\ldots,|\mathcal{D}|$ is powered by the harvested energy during its prior slot $d_i$ (c.f. Proposition \ref{proposition:Ab}). The required amount of energy for information transmission is harvested at an earlier slot to ensure that there is always sufficient energy for WIT, i.e., no energy outage at the energy buffer. 
Motivated by this observation, we partition the transmission block $\mathcal{K}$ into subsets, referred to as {\it windows}, denoted by $\mathcal{W}_i,i=1,\ldots,W$, where $W$ denotes the number of windows. WET is performed in each window $\mathcal{W}_i,i=1,\ldots,W-1$, and the harvested energy during $\mathcal{W}_i$ is utilized to power information transmission during the next window $\mathcal{W}_{i+1}$, which ensures no energy outage for WIT during the block (except the first window). No WET is performed in the last window. In particular, the first window consists of the first slot, while the remaining $K-1$ slots are partitioned into $W-1$ windows, each window consists of $L$ slots, where $L$ denotes the window size, with $1\leq L\leq K-1$. For simplicity, we assume $K-1$ is divisible by $L$; hence, $W=(K-1)/L+1$. Notice that the partitioned windows for the causal CSI case are fixed, which is independent of the channels for the WET link.
                                  
In each window $\mathcal{W}_i,i=1,\ldots,W-1$, one SC is selected to perform WET. We assume the transmission energy at EAP is equally scheduled to the windows $\mathcal{W}_i,i=1,\ldots,W-1$, hence, the EAP transmit with power $KQ/(W-1)$ at the selected SC in each window.
For information transmission at the user, two energy sources are available, i.e., the initial battery energy $B_1$ and the energy harvested from EAP. Since only causal CSI is available, we assume $B_1$ is equally scheduled for information transmission over all $K$ slots, hence each slot is scheduled with transmission power $B_1/K$. The energy harvested during window $\mathcal{W}_i,i=1,\ldots,W-1,$ is utilized for information transmission during the next window $\mathcal{W}_{i+1}$, where each slot is scheduled with equal transmission power. At each slot $k$, $\{p_{k,n},n\in\Seti\}$ is obtained by the WF power allocation\footnote{In practice, the user may be imposed on a peak power constraint on its transmission power on each SC $n$ during each slot $k$, i.e., $p_{k,n}\leq P_{\rm peak}$. In this case, at each slot $k$, the power allocation at the user $\{p_{k,n},n\in\Seti\}$ is then obtained by the (revised) WF power allocation with additional peak power constraint $p_{k,n}\leq P_{\rm peak}$.} over SCs $n\in\Seti$.

Next, we investigate the SC selection in each window $\mathcal{W}_i,i=1,\ldots,W-1$. As revealed by the full CSI case, WET is performed on one SC to power its subsequent interval, hence, we aim to select one SC that is expected to have the largest channel power gain for the WET link among all SCs in each window to perform WET. It is necessary for a SC to be best among all SCs in a window that it is the best SC at its current slot, hence, the SC is selected from the set $\{\tilde{\Pi}(k),k\in\mathcal{W}_i\}$. For the first window $\mathcal{W}_1=\{1\}$, the best SC $\tilde{\Pi}(1)$ is selected to perform WET. Consider other windows $\mathcal{W}_i,i=2,\ldots,W-1$.
Assume the channel power gain at the best SC at the $k$th slot in the window is denoted by $h_{[k]}$, where $k=1,\ldots,L$. The SC selection problem is then formulated as a stopping problem described as follows. Given a sequentially occurring random sequence $h_{[1]},h_{[2]},\ldots,h_{[L]}$, the permutations of which are equally likely, our objective is to select a slot to stop, the index of which is denoted by $s$, such that the probability of $h_{[s]}>h_{[j]},\forall j=1,\ldots,L,j\neq s$, denoted by $P_{\rm r}$, is maximized. 
The challenge is that at any slot $k=1,\ldots,L$, the decision of whether to stop at current slot (i.e., $s=k$) or stop at latter slots (i.e., $s\neq k$) needs to be made immediately, based on causal information, i.e., $\{h_{[j]},1\leq j\leq k\}$. The decision of $s=k$ suffers a potential loss when better channels occur in subsequent slots in the window; whereas the decision of $s\neq k$ risks the probability that a better channel never occurs subsequently. The stopping problem can be viewed as a classic Secretary Problem \cite{Ferguson}. 
A necessary condition for stopping at slot $s$ is that $h_{[s]}>h_{[j]},\forall j=1,\ldots,L,j< s$, i.e., slot $s$ causally dominates all its previous slots in the window; otherwise, the probability $P_{\rm r}$ becomes zero. Hence, the optimal stopping rule lies in a class of policies, which are described as follows: Define the cutoff slot $f(L)$, which is a parameter that can be optimized, and $1\leq f(L)\leq L$. The first $f(L)-1$ slots are for observation. During the remaining $L-f(L)+1$ slots, the first slot (if any) that causally dominates all its previous slots, is selected as $s$; if no slot is selected until the last slot, then $s=L$. The probability $P_{\rm r}$ is given by \cite{Ferguson}
\begin{align}\label{eq:probability}
P_{\rm r}=
\begin{cases}
\frac{1}{L}, & f(L)=1,\\
\frac{f(L)-1}{L}\sum\limits_{l=f(L)}^L\frac{1}{l-1}, & 1<f(L)\leq L.
\end{cases}
\end{align}
The optimal cutoff slot $f^\ast(L)$ that maximizes $P_{\rm r}$ is thus obtained as $f^\ast(L)=\underset{1\leq f(L)\leq L}{\arg\max}~P_{\rm r}$. The above SC selection scheme is referred to as {\it dynamic SC with observe-then-transmit (OTT)}.

\begin{figure}
\centering
\includegraphics[width=8cm]{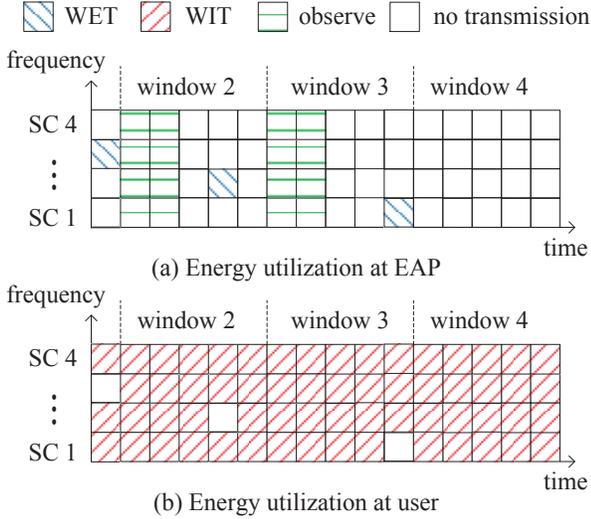}  
\caption{Energy utilization for the scheme of dynamic SC with OTT, where $K=16,N=4$ and $L=5$.}
\label{fig:window}
\end{figure}

An example of the scheme of dynamic SC with OTT is illustrated in Fig. \ref{fig:window}, where the total number of slots is $K=16$, and the window size is $L=5$. The windows are obtained as $\mathcal{W}_1=\{1\},\mathcal{W}_2=\{2,\ldots,6\},\mathcal{W}_3=\{7,\ldots,11\},\mathcal{W}_4=\{12,\ldots,16\}$. In $\mathcal{W}_1$, the best SC (SC 3) at WET link in slot one is selected to transmit energy. From \eqref{eq:probability}, the cutoff slot is obtained as $f^\ast(L)=3$. Hence, for $\mathcal{W}_2$ and $\mathcal{W}_3$, in each window the first two slots are for observing $h_{k,n},n\in\mathcal{N}$, and the first slot (if any) during the remaining three slots that causally dominates all its previous slots in the window is selected for WET; otherwise, the last slot is scheduled for WET. In $\mathcal{W}_4$, there is no energy transmission from EAP. In Fig. \ref{fig:window}(a), WET is performed at SC 3 during slot 1, SC 2 during slot 5, and SC 1 during slot 11; hence, in Fig. \ref{fig:window}(b), the user transmits information at the remaining SCs.

\section{Numerical Example}\label{sec:numerical}

In this section, we provide numerical examples. We focus on the separated EAP/DAP case, where the distances from EAP to the user and from the user to DAP are assumed to be 3meter (m) and 7m, respectively. The total number of slots is set to be $K=61$. The total bandwidth of the system is assumed to be $64$MHz, centered at $900$MHz, which is equally divided into $N=16$ SCs, each with bandwidth $4$MHz.
The frequency-selective channels are generated by a multi-path power delay profile with exponential distribution $A(\tau)=\frac{1}{\sigma_{\rm rms}}e^{-\tau/\sigma_{\rm rms}},\tau>0$, where $\sigma_{\rm rms}$ denotes the root mean square (rms) delay spread. Assume $\sigma_{\rm rms}=0.02\mu$s, the coherence bandwidth is therefore given by $B_{\rm c}=\frac{1}{2\pi\sigma_{\rm rms}}\approx 8$MHz. The channels over slots are generated independently. In later simulations, all achievable rates are averaged over $10^4$ independent channel realizations. 
Assuming the path-loss exponent is three, the signal power attenuation at transmission distance $d$ (in meter) is then approximately $(-31.5-30\log_{10}d)$dB \cite{Goldsmith-wireless}. The receiver noise power spectrum density at DAP is assumed to be $-174$dBm/Hz, and $\Gamma=9$dB. The initial batter energy is set to be $B_1=0$. The energy efficiency $\zeta$ is assumed to be 0.2.

\subsection{Offline Algorithms Under Full CSI}
First, consider the full CSI case, in which we compare the performance by different offline schemes. As benchmark, we consider the system in \cite{Ho,Ozel} with random energy arrivals at the EH user. In particular, the EAP in Fig. \ref{fig:system model} is replaced by an ambient RF transmitter
which is oblivious of the WET link, and hence its transmit power over time is random to the user, since it is adapted to its own information transmission link (to another receiver).
Throughout the whole transmission block, the ambient transmitter transmits over a fixed SC (e.g., the first SC), and the remaining $(N-1)$ SCs are for the information transmission at the EH user.
In simulation, the transmit power at the ambient transmitter $q_{k,1},k\in\mathcal{K}$ are randomly generated by the uniform distribution over $[0, 1]$, and then are normalized such that $1/K\sum_{k=1}^K q_{k,1}=Q$. Hence, during each slot $k,k\in\mathcal{K}$, a random energy $h_{k,1}q_{k,1}$ arrives at the user. Given $\{h_{k,1}q_{k,1},k\in\mathcal{K}\}$, the achievable rates are obtained by optimizing $\{p_{k,n},k\in\mathcal{K},n\in\mathcal{N}\backslash\{1\}\}$ according to \cite{Ho,Ozel}. The performance of this system is obtained by averaging the results from $10^4$ realizations of random transmission power $\{q_{k,1},k\in\mathcal{K}\}$. In addition, the performance upper bound obtained by the ideal DAP with perfect interference cancellation (refer to Remark \ref{remark:upper bound}) is also considered as benchmark. Besides the optimal joint WET/WIT transmission, for comparison we also consider a sub-optimal WET scheme referred to as constant WET, where the EAP transmits constant power $Q$ each slot at given SC (by dynamic/static SC schemes).

\begin{figure}
\centering
\includegraphics[width=9cm]{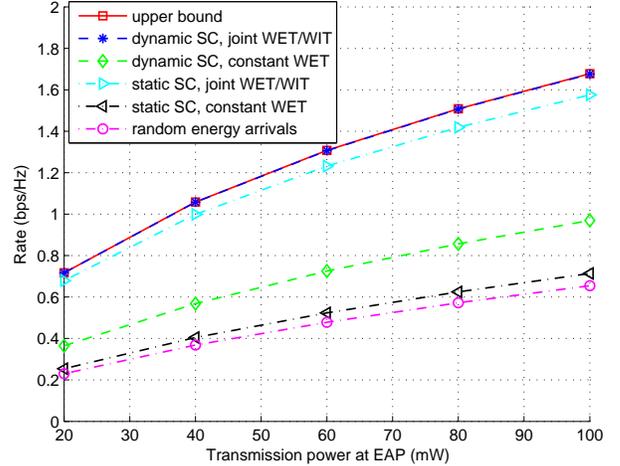}  
\caption{Performance comparison for offline algorithms when full CSI is available.}
\label{fig:compare offline} 
\end{figure}

Fig. \ref{fig:compare offline} shows the achievable rates at DAP versus transmission power at EAP by different offline schemes. In Fig. \ref{fig:compare offline}, it is observed that the achievable rates by the proposed dynamic SC scheme with joint WET/WIT transmission are very close to that by the upper bound. 
Comparing the joint WET/WIT with constant WET transmission schemes, it is observed that for both dynamic and static SC, the joint WET/WIT schemes achieve much larger rates than that by the constant WET schemes, which demonstrates the importance of optimal energy allocation over time for the WET link.
Comparing the dynamic SC and the static SC schemes, it is observed that for either joint WET/WIT or constant WET transmission, the dynamic SC scheme is superior than the static SC scheme, and the performance gap is larger when EAP performs constant WET transmission. This is because that the dynamic SC scheme exploits more frequency diversity for WET; in contrast, the available channels for WET are constrained on one SC over the whole transmission block. Hence, in general more energy can be harvested to support higher data rate by the dynamic SC scheme than by the static SC scheme. It implies that optimizing SC allocation is important to the performance, especially when EAP performs sub-optimal constant-power WET.
Last, comparing the achievable rates by the wireless powered communication system with dynamic SC, joint WET/WIT and that by the system with random energy arrivals, a remarkable performance improvement is observed by the wireless powered system, which demonstrates the superiority of WPCN with dedicated EAP over conventional EH system with random energy arrivals.

\begin{figure}
\centering
\includegraphics[width=9cm]{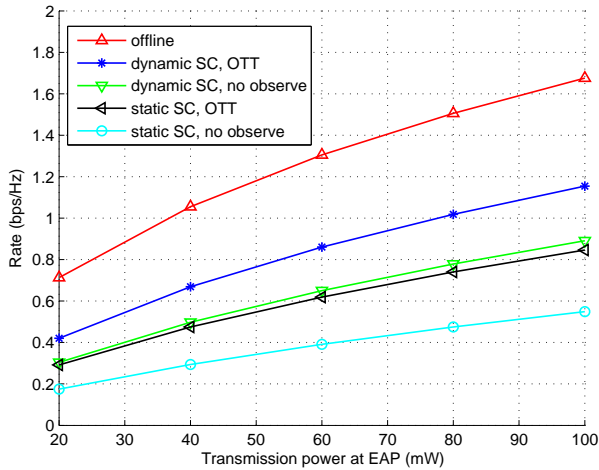} 
\caption{Performance comparison for online algorithms when casual CSI is available, where $L=15$.}
\label{fig:compare online Q} 
\end{figure}

\begin{figure}
\centering
\includegraphics[width=9cm]{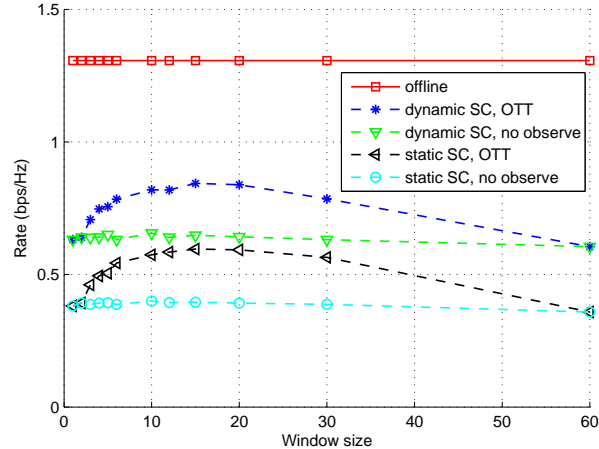} 
\caption{Performance comparison for online algorithms when casual CSI is available, where $Q=60$mW.}
\label{fig:compare online L} 
\end{figure}

\subsection{Online Algorithms Under Causal CSI}
Next, consider the causal CSI case, in which we compare the performance by different online schemes. Besides the scheme of dynamic SC with OTT proposed in Section \ref{sec:causal csi}, for comparison we also consider other window-based online schemes, where a static SC (e.g., the first SC) is fixed for WET, or the EAP selects the first slot in each window (i.e., no channel observation) to perform WET.
In addition, the performance by the offline dynamic SC with joint WET/WIT scheme is considered as a benchmark.

Fig. \ref{fig:compare online Q} shows the achievable rates by different window-based online schemes versus the transmission power at EAP $Q$. In Fig. \ref{fig:compare online Q}, the window size $L$ is set to be $L=15$ with optimal cutoff slot $f^\ast(L)=6$. 
It is observed in Fig. \ref{fig:compare online Q} that the achievable rates by online schemes are smaller than that by the offline scheme, due to lack of future information of channels for the WET/WIT links. 
In Fig. \ref{fig:compare online Q}, comparing the performance by the dynamic and static SC schemes, it is observed that the dynamic SC schemes achieve larger rates. Comparing the performance by the OTT schemes and that by the no-observation schemes, it is observed that the OTT schemes are superior, as observation for the WET link helps to employ more efficient energy transmission by transmitting on SC that is expected to have large channel power gain.

Fig. \ref{fig:compare online L} further shows the achievable rates by different window-based online schemes versus the window size $L$. In Fig. \ref{fig:compare online L}, the transmission power at EAP is set to be $Q=60$mW. Similar as in Fig. \ref{fig:compare online Q}, it is observed in Fig. \ref{fig:compare online L} that the dynamic SC scheme is superior than the static SC scheme, and the OTT scheme is better than the no-observation scheme. We notice that in Fig. \ref{fig:compare online L} the performance by the OTT schemes degrade to that by the no-observation schemes when $L=1,2,60$, as no observation is performed for these special cases. Furthermore, in Fig. \ref{fig:compare online L}, it is observed that as the window size increases, the achievable rates by the no-observation schemes are independent of the window size; whereas the achievable rates by the OTT schemes first increase and then decrease. Intuitively, this may be because that with larger window size more observation slots help to select SCs with large channel power gain to perform WET. However, smaller window size results in more number of selected SCs, which helps to compensate the loss of selecting poor SCs (in the last slot of each window).

\section{Conclusion}
This paper studied an OFDM-based wireless powered communication system, where a user harvests energy from the EAP to power its information transmission to the DAP. 
The energy transmission by the EAP and the information transmission by the user is performed over orthogonal SCs. The achievable rate at the DAP is maximized by jointly optimizing the SC allocation over time and power allocation over time and SCs for both WET and WIT links. Numerical results demonstrate that by dynamic SC allocation and joint power allocation, the performance is improved remarkably as compared to a conventional EH system where the information transmitter is powered by random energy arrivals. Furthermore, both cases of full CSI and causal CSI are considered. Numerical results show that for the case of causal CSI utilizing partial information of the channels for the WET link can be beneficial to the achievable rate. In general, there is still performance gap between the online and offline algorithms. Therefore, how to further reduce this gap is worthy of future investigation. In this paper we consider the single-user scenario for the purpose of exposition. As an extension, it is interesting to investigate the more general scenario of multiple users co-existing in the system. For the multiuser case, the broadcast signals by the EAP may provide wireless power to all users simultaneously. The information transmission from different users to the DAP may be coordinated by the orthogonal frequency division multiple access (OFDMA) scheme. Hence, at each slot, besides the SC allocation between WET and WIT links, the SCs for the WIT links need to be further allocated among multiple users. The complicated twofold SC allocation inevitably results in a non-convex optimization problem, which in general is challenging to solve optimally.

\appendices
\section{Proof of Proposition \ref{proposition:A}}\label{appendix:proof proposition A}
Since no WET is performed at the last slot $K$, we prove Proposition \ref{proposition:A} for $1\leq k\leq K-1$. For optimal $q_{k,n}^\ast,k\in\mathcal{K},n\in\Sete$, assume there exists a slot $j,1\leq j\leq K-1$ and SC $l\in\mathcal{N}_j^{\rm E},l\neq m(j)$, such that $q_{j,l}^\ast>0$.
We construct a different power allocation for the WET link as follows: 
\begin{align}\label{eq:proof 1 q}
\hat{q}_{k,n}=\begin{cases}
\sum\limits_{u\in\Sete} q_{j,u}^\ast, & k=j,n=m(j),\\
0, & k=j,n\neq m(j),\\
q_{k,n}^\ast, & k\neq j, n\in\Sete.
\end{cases}
\end{align}
From \eqref{eq:proof 1 q}, $\hat{q}_{k,n},k\in\mathcal{K},n\in\Sete$ satisfies \eqref{eq:full CSI c1}. Since $q_{j,l}^\ast>0$, we have
\begin{equation}\label{eq:proof 1 larger}
\sum\limits_{n\in\mathcal{N}_j^{\rm E}} h_{j,n}(\hat{q}_{j,n}-q_{j,n}^\ast)=
\sum\limits_{n\in\mathcal{N}_j^{\rm E}} \left(h_{j,m(j)}-h_{j,n}\right)q_{j,n}^\ast>0.
\end{equation}
From \eqref{eq:proof 1 larger}, by $\hat{q}_{k,n},k\in\mathcal{K},n\in\Sete$, a larger feasible region for $p_{k,n},k\in\mathcal{K},n\in\Seti$ is obtained than that by $q_{k,n}^\ast,k\in\mathcal{K},n\in\Sete$, thus a larger achievable rate can be obtained by increasing some $p_{K,n},n\in\mathcal{N}$, which contradicts the assumption that $q_{k,n}^\ast,k\in\mathcal{K},n\in\Sete$ is optimal. Hence, $q_{k,n}^\ast=0$ for $n\neq m(k)$. Proposition \ref{proposition:A} is thus proved.

\section{Proof of Proposition \ref{proposition:Aa}}\label{appendix:proof proposition Aa}
Given SC allocation $\Pi(k),k\in\mathcal{K}$, there are two possible cases for slots in set $\mathcal{D}^{\rm c}$, i.e., $\Pi(k)=0$ or $\Pi(k)\in\mathcal{N}$.
For $k\in\mathcal{D}^{\rm c},\Pi(k)=0$, we have $q_k^\ast=0$, since no SC is available for WET during the slot $k$. Next, we prove that $q_k^\ast=0$ for $k\in\mathcal{D}^{\rm c},\Pi(k)\in\mathcal{N}$.
For any power allocation $\{q_k\}$, $\{p_{k,n}\}$ that satisfy the constraints (\ref{eq:full CSI re1 c1}) and (\ref{eq:full CSI re1 c2}), assume there exists a slot $i\in\mathcal{D}^{\rm c}$ with $\Pi(i)\in\mathcal{N}$ and $q_i>0$. By the definition of set $\mathcal{D}$, there exists a slot $1\leq j<i$ such that $h_{j}> h_{i}>0,\Pi(j)\in\mathcal{N}$. We construct a power allocation strategy $\{\hat{q}_k\}$, $\{\hat{p}_{k,n}\}$ given by
\begin{align}
\hat{q}_k=\begin{cases}
q_j+q_i, & k=j,\\
0, & k=i,\\
q_k, & {\rm otherwise}.\nonumber
\end{cases}
\end{align}
\begin{align}
\hat{p}_{k,n}=\begin{cases}
p_{k,n}+\frac{\zeta (h_{j}-h_{i})q_i}{N}, & k=i,n\in\Seti,\\
p_{k,n}, & {\rm otherwise}.\nonumber
\end{cases}
\end{align}
It can be verified that $\{\hat{q}_k\}$ and $\{\hat{p}_{k,n}\}$ satisfy the constraints (\ref{eq:full CSI re1 c1}) and (\ref{eq:full CSI re1 c2}). Since $h_{j}> h_{i}$ and $q_i>0$, the achievable rate by $\{\hat{q}_k\}$, $\{\hat{p}_{k,n}\}$ is larger than that by $\{q_k\}$, $\{p_{k,n}\}$, i.e., $\{q_k\}$, $\{p_{k,n}\}$ is not optimal. Hence, the optimal solution satisfies that $q_k^\ast=0$ for $k\in\mathcal{D}^{\rm c}$. The proof of Proposition \ref{proposition:Aa} is thus completed.

\section{Proof of Proposition \ref{proposition:Ab}}\label{appendix:proof proposition Ab}
We prove that if $\{q_k^\ast\}$ and $\{p_{k,n}^\ast\}$ satisfy (\ref{eq:full CSI re1 c2}) with equality at $i=d_j$, where $1\leq j\leq |\mathcal{D}|$, then they satisfy (\ref{eq:full CSI re1 c2}) with equality at $i=d_{j+1}$, i.e.,
\begin{equation}\label{eq:proof dl}
\sum\limits_{k=1}^{d_{j+1}}\sum\limits_{n\in\Seti}p_{k,n}^\ast=\zeta\sum\limits_{k=1}^{d_{j+1}-1}h_{k}q_k^\ast+B_1.
\end{equation}
Note that (\ref{eq:proof dl}) is satisfied for $j=|\mathcal{D}|$; otherwise, the objective function in Problem (\ref{prob:full CSI re1}) can be increased by increasing some $p_{K,n}$. 

Next, we prove (\ref{eq:proof dl}) for the case $1\leq j\leq |\mathcal{D}|-1$ by contradiction.
The optimal solutions $\{q_k^\ast\}$ and $\{p_{k,n}^\ast\}$ satisfy the constraints (\ref{eq:full CSI re1 c1}) and (\ref{eq:full CSI re1 c2}).
Assume $\{q_k^\ast\}$ and $\{p_{k,n}^\ast\}$ do not satisfy (\ref{eq:proof dl}), i.e., $\Delta\triangleq \zeta\sum_{k=1}^{d_{j+1}-1}h_{k}q_k^\ast+B_1-\sum_{k=1}^{d_{j+1}}\sum_{n\in\Seti}p_{k,n}^\ast>0$. From (\ref{eq: first equal A}) and Proposition \ref{proposition:Aa}, we have
\begin{equation}\label{eq:delta}
\Delta=\zeta h_{d_j}q_{d_j}^\ast-\sum\limits_{k=d_j+1}^{d_{j+1}}\sum\limits_{n\in\Seti}p_{k,n}^\ast.
\end{equation}
Now, we construct a power allocation strategy $\{\hat{q}_k,k\in\mathcal{K}\}$, $\{\hat{p}_{k,n},k\in\mathcal{K},n\in\Seti\}$ given by
\begin{equation}\label{eq:qi hat}
\hat{q}_k=\begin{cases}
q_{d_j}^\ast-\frac{\Delta}{\zeta h_{d_j}}, & k=d_j, \\
q_{d_{j+1}}^\ast+\frac{\Delta}{\zeta h_{d_j}}, & k=d_{j+1}, \\
q_k^\ast, & {\rm otherwise}.
\end{cases}
\end{equation}
\begin{equation}\label{eq:pi hat}
\hat{p}_{k,n}=\begin{cases}
p_{k,n}^\ast, & k=1,\ldots,d_{j+1},  \\
p_{k,n}^\ast+\frac{\left(h_{d_{j+1}}-h_{d_j}\right)\Delta}{h_{d_j}(K-d_{j+1})N}, & k=d_{j+1}+1,\ldots,K.
     \end{cases}
\end{equation}
It can be verified that $\{\hat{q}_k\}$ and $\{\hat{p}_{k,n}\}$ satisfy the constraints (\ref{eq:full CSI re1 c1}) and (\ref{eq:full CSI re1 c2}). Since $\Delta>0$ and $h_{d_{j+1}}>h_{d_j}$, the power allocation $\{\hat{q}_k\}$ and $\{\hat{p}_{k,n}\}$ achieve larger rate than $\{q_k^\ast\}$ and $\{p_{k,n}^\ast\}$, which contradicts the assumption that $\{q_k^\ast\}$ and $\{p_{k,n}^\ast\}$ are optimal for \eqref{prob:full CSI re1}. 
Therefore, $\{q_k^\ast\}$ and $\{p_{k,n}^\ast\}$ satisfy (\ref{eq:proof dl}). By induction, 
\begin{equation}\label{eq:proof induct}
\sum\limits_{k=1}^{d_{l+1}}\sum\limits_{n\in\Seti}p_{k,n}^\ast=\zeta \sum\limits_{k=1}^{d_{l+1}-1}h_{k}q_k^\ast+B_1, ~~ l=j,\ldots,|\mathcal{D}|.
\end{equation}
It follows from (\ref{eq:proof induct}) that, for $l=j,\ldots,|\mathcal{D}|$,
\begin{equation}
\sum\limits_{k\in\mathcal{D}_{l+1}}\sum\limits_{n\in\Seti} p_{k,n}^\ast =\zeta \sum\limits_{k=d_l}^{d_{l+1}-1}h_{k}q_k^\ast= h_{d_{l}}q_{d_{l}}^\ast
\end{equation}
which completes the proof of Proposition \ref{proposition:Ab}.

\section{Proof of Lemma \ref{lemma:Aa}}\label{appendix:proof lemma Aa}
For the case $1\leq x\leq 2$, from (\ref{eq:C2-dx A}), \eqref{eq:lemma Aa} is satisfied.
For the case $2<x\leq |\mathcal{D}|+1$, we first prove $q_{d_k}^\ast=0$ for $1\leq k\leq x-2$ by contradiction. Assume there exists $q_{d_j}^\ast>0$ for $1\leq j\leq x-2$.
Define
$\Delta\triangleq\min\limits_{i=j+1,\ldots,x-1}\left(\zeta\sum\limits_{k=1}^{i-1}h_{d_k}q_{d_k}+B_1-\sum\limits_{k=1}^{d_i}\sum\limits_{n\in\Seti}p_{k,n}\right)$. 
From (\ref{eq:C2-less A}), we have $\Delta>0$.
We construct a power allocation strategy $\{\hat{q}_k,k\in\mathcal{K}\}$ and $\{\hat{p}_{k,n},k\in\mathcal{K},n\in\Seti\}$ given by
\begin{align}
\hat{q}_k=\begin{cases}
q_{d_j}^\ast-\min\left(q_{d_j}^\ast,\frac{\Delta}{\zeta h_{d_j}}\right), & k=d_j,\nonumber\\
q_{d_{x-1}}^\ast+\min\left(q_{d_j}^\ast,\frac{\Delta}{\zeta h_{d_j}}\right), & k=d_{x-1},\nonumber\\
q_k^\ast, & {\rm otherwise}.
\end{cases}
\end{align}
\begin{align}
\hat{p}_{k,n}&=\begin{cases}
p_{k,n}^\ast+\frac{\zeta (h_{d_{x-1}}-h_{d_j})}{N}\min\left(q_{d_j}^\ast,\frac{\Delta}{\zeta h_{d_j}}\right),  
& k=d_x,\nonumber\\
p_{k,n}^\ast, & {\rm otherwise}. \nonumber
\end{cases}
\end{align}
It can be verified that $\{\hat{q}_k\}$ and $\{\hat{p}_{k,n}\}$ satisfy the constraints (\ref{eq:full CSI re1 c1}) and (\ref{eq:full CSI re1 c2}). Since $h_{d_{x-1}}>h_{d_j}$, $q_{d_j}^\ast>0$, and $\Delta>0$, the power allocation $\{\hat{q}_k\}$ and $\{\hat{p}_{k,n}\}$ achieve larger rate than $\{q_k^\ast\}$ and $\{p_{k,n}^\ast\}$, which contradicts the assumption that $\{q_k^\ast\}$ and $\{p_{k,n}^\ast\}$ are optimal for \eqref{prob:full CSI re1}. Therefore, $q_{d_k}^\ast=0$ for $1\leq k\leq x-2$. Then \eqref{eq:lemma Aa} follows from (\ref{eq:C2-dx A}). The proof of Lemma \ref{lemma:Aa} then completes.

\section{Proof of Lemma \ref{lemma:Ab}}\label{appendix:proof lemma Ab}
Given $\{\Pi(k)\}$, we prove the equivalence between Problems \eqref{prob:full CSI re1} and \eqref{prob:full CSI re2}. It is sufficient for us to prove that given optimal solution $\{q_{k,n}\}$, $\{p_{k,n}\}$ for Problem \eqref{prob:full CSI re1}, $\{p_{k,n}'\}$ obtained by \eqref{eq:pi'-caseI A} is optimal for Problem \eqref{prob:full CSI re2}; given optimal solution $\{p_{k,n}'\}$ for Problem \eqref{prob:full CSI re2}, $\{q_k\}$, $\{p_{k,n}\}$ obtained by \eqref{eq:qi opt A} and \eqref{eq:pi'-caseI A} is optimal for Problem \eqref{prob:full CSI re1}.
For convenience, the optimal value of Problems \eqref{prob:full CSI re1} and \eqref{prob:full CSI re2} are denoted by $R^\ast$ and $R'$, respectively. 

Given optimal solution $\{q_{k,n}\}$, $\{p_{k,n}\}$ for Problem \eqref{prob:full CSI re1}, then
$\{q_{k,n}\}$, $\{p_{k,n}\}$ satisfy constraints \eqref{eq:full CSI re1 c1} and \eqref{eq:full CSI re1 c2}. We obtain $\{p_{k,n}'\}$ by \eqref{eq:pi'-caseI A}. Since $g_{k,n}p_{k,n}=g_{k,n}'p_{k,n}'$, the average rate achieved by $\{p_{k,n}'\}$ equals to $R^\ast$. 
Next, we prove that $\{p_{k,n}'\}$ is a feasible solution for Problem \eqref{prob:full CSI re2}.
From Lemma \ref{lemma:Aa}, \eqref{eq:pi'-caseI A}, and \eqref{eq:C2-less A}, $\{p_{k,n}'\}$ satisfy constraints \eqref{eq:full CSI re2 c2} and \eqref{eq:full CSI re2 c3}. From Proposition \eqref{eq:C2-dx A}, and \ref{proposition:Ab}, $\{q_{k,n}\}$, $\{p_{k,n}\}$ satisfy 
\begin{equation}
\sum\limits_{k\in\mathcal{D}_{l+1}}\sum\limits_{n\in\Seti} p_{k,n}^\ast = \zeta h_{d_{l}}q_{d_{l}}^\ast, ~~ l=x,\ldots,|\mathcal{D}|. \label{eq:proof lemma b}
\end{equation}
From Proposition \ref{proposition:Aa}, \eqref{eq:lemma Aa}, and \eqref{eq:pi'-caseI A}, it follows that
\begin{equation}
\sum\limits_{k=1}^{K}\sum\limits_{n\in\Seti} p_{k,n}'\leq \zeta\sum\limits_{i=x-1}^{|\mathcal{D}|}q_{d_i}+\frac{B_1}{h_{d_{x-1}}} 
\leq \zeta KQ+\frac{B_1}{h_{d_{x-1}}}.
\end{equation}
It follows that $\{p_{k,n}'\}$ satisfy constraint \eqref{eq:full CSI re2 c1}; thus, $\{p_{k,n}'\}$ is a feasible solution for Problem \eqref{prob:full CSI re1}. Therefore, the average rate achieved by $\{p_{k,n}'\}$ is no larger than $R'$; i.e., $R^\ast\leq R'$, where the equality holds if and only if $\{p_{k,n}'\}$ is optimal for Problem \eqref{prob:full CSI re2}. 

Given optimal solution $\{p_{k,n}'\}$ for Problem \eqref{prob:full CSI re2}, then $\{p_{k,n}'\}$ satisfy constraints \eqref{eq:full CSI re2 c1}, \eqref{eq:full CSI re2 c2}, and \eqref{eq:full CSI re2 c3}. 
We obtain $q_k$ and $p_{k,n}$ by \eqref{eq:qi opt A} and \eqref{eq:pi'-caseI A}. Since $g_{k,n}p_{k,n}=g_{k,n}'p_{k,n}'$, the average rate achieved by $\{q_k\}$, $\{p_{k,n}\}$ equals to $R'$.
Next, we prove that $\{q_k\}$, $\{p_{k,n}\}$ is a feasible solution for Problem \eqref{prob:full CSI re1}.
From \eqref{eq:full CSI re2 c1} and \eqref{eq:qi opt A}, $\{q_k\}$ satisfy constraint \eqref{eq:full CSI re1 c1}. From \eqref{eq:pi'-caseI A}, \eqref{eq:full CSI re2 c2}, and \eqref{eq:qi opt A}, $\{q_k\}$ and $\{p_{k,n}\}$ satisfy constraints \eqref{eq:full CSI re1 c1} and \eqref{eq:full CSI re1 c1}. Therefore, $\{q_k\}$, $\{p_{k,n}\}$ is a feasible solution for Problem \eqref{prob:full CSI re1}. It follows that the average rate achieved by $\{q_k\}$, $\{p_{k,n}\}$ is no larger than $R^\ast$; thus, $R'\leq R^\ast$, where the equality holds if and only if $\{q_k\}$, $\{p_{k,n}\}$ is optimal for Problem \eqref{prob:full CSI re1}. 

From $R^\ast\leq R'$ and $R'\leq R^\ast$, we have $R^\ast=R'$. Therefore, given optimal solution $\{q_{k,n}\}$, $\{p_{k,n}\}$ for Problem \eqref{prob:full CSI re1}, $\{p_{k,n}'\}$ obtained by \eqref{eq:pi'-caseI A} is optimal for Problem \eqref{prob:full CSI re2}; given optimal solution $\{p_{k,n}'\}$ for Problem \eqref{prob:full CSI re2}, $\{q_k\}$, $\{p_{k,n}\}$ obtained by \eqref{eq:qi opt A} and \eqref{eq:pi'-caseI A} is optimal for Problem \eqref{prob:full CSI re1}. The proof of Lemma \ref{lemma:Ab} completes.

\section{Proof of Proposition \ref{proposition:Ac}}\label{appendix:proof proposition Ac}
Problem (\ref{prob:full CSI re2}) is a convex optimization problem, and thus can be
optimally solved by applying the Lagrange duality method. The Lagrangian of Problem (\ref{prob:full CSI re2}) is given by
\begin{align}
\mathcal{L}\left(\{p_{k,n}'\},\lambda,\delta,\mu\right)=& \frac{1}{KN}\sum\limits_{k=1}^{K}\sum\limits_{n\in\Seti}\log_2\left(1+\frac{g_{k,n}'p_{k,n}'}{\N}\right) \nonumber\\
& +\lambda\left(\zeta KQ+\frac{B_1}{h_{d_{x-1}}}-\sum\limits_{k=1}^{K}\sum\limits_{n\in\Seti} p_{k,n}'\right)\nonumber\\
& +\delta\left(\frac{B_1}{h_{d_{x-1}}}-\sum\limits_{k=1}^{d_{x-1}}\sum\limits_{n\in\Seti}p_{k,n}'\right) \nonumber\\
& +\mu\left(\sum\limits_{k=1}^{d_x}\sum\limits_{n\in\Seti}p_{k,n}'- \frac{B_1}{h_{d_{x-1}}}\right)
\end{align}
where $\lambda,\delta$, and $\mu$ are the non-negative dual variables associated with the corresponding constraints in Problem (\ref{prob:full CSI re2}). The necessary and sufficient conditions for $\{p_{k,n}'\}$ and $\lambda,\delta,\mu$ to be both primal and dual optimal are given by the Karush-Kuhn-Tucker (KKT) optimality conditions: $\{p_{k,n}'\}$ satisfy all the constraints in Problem (\ref{prob:full CSI re2}), and
\begin{subequations}
\begin{align}
\lambda\left(\zeta KQ+\frac{B_1}{h_{d_{x-1}}}-\sum\limits_{k=1}^{K}\sum\limits_{n\in\Seti} p_{k,n}'\right)&=0, \tag{\theequation a}\label{eq:KKT a}\\
\delta\left(\frac{B_1}{h_{d_{x-1}}}-\sum\limits_{k=1}^{d_{x-1}}\sum\limits_{n\in\Seti}p_{k,n}'\right)&=0,
\tag{\theequation b}\label{eq:KKT b}\\
\mu\left(\sum\limits_{k=1}^{d_x}\sum\limits_{n\in\Seti}p_{k,n}'- \frac{B_1}{h_{d_{x-1}}}\right)&=0,
\tag{\theequation c}\label{eq:KKT c}\\
\frac{\partial\mathcal{L}\left(\{p_{k,n}'\},\lambda,\delta,\mu\right)}{\partial p_{k,n}'}&=0.
\tag{\theequation d}\label{eq:KKT d}
\end{align}
\end{subequations}
From (\ref{eq:C2-less A}) and Lemma \ref{lemma:Aa}, $\sum_{k=1}^{d_{x-1}}\sum_{n\in\Seti}p_{k,n}^\ast<B_1$; therefore, the optimal $\{p_{k,n}'\}$ satisfies $\sum_{k=1}^{d_{x-1}}\sum_{n\in\Seti}p_{k,n}'$ $<{B_1}/{h_{d_{x-1}}}$. It follows that the optimal $\delta=0$ by (\ref{eq:KKT b}). From (\ref{eq:KKT d}) and $\delta=0$, the optimal $p_{k,n}',k\in\mathcal{K},n\in\Seti$ is given by
\begin{align}
p_{k,n}'=\begin{cases}
\left(\frac{1}{(\lambda-\mu)KN\ln 2}-\frac{\N}{g_{k,n}'}\right)^+, & k = 1,\ldots,d_{x}, \\
\left(\frac{1}{\lambda KN\ln 2}-\frac{\N}{g_{k,n}'}\right)^+, & k=d_x+1,\ldots,K.
\end{cases}
\end{align}
If the optimal $\mu>0$, from (\ref{eq:KKT a}) and (\ref{eq:KKT c}), we have $\sum_{k=1}^{d_x}\sum_{n\in\Seti}p_{k,n}'={B_1}/{h_{d_{x-1}}}$ and $\sum_{k=d_x+1}^{K}\sum_{n\in\Seti}p_{k,n}'$ $=\zeta KQ$.
If the optimal $\mu=0$, then
\begin{align}
p_{k,n}'=\left(\frac{1}{\lambda KN\ln 2}-\frac{\N}{g_{k,n}'}\right)^+,  k\in\mathcal{K},n\in\Seti
\end{align}
where $\lambda$ satisfies $\sum_{k=1}^K\sum_{n\in\Seti} p_{k,n}'=\zeta KQ+{B_1}/{h_{d_{x-1}}}$ by (\ref{eq:KKT a}). Proposition \ref{proposition:Ac} is thus proved.

\section{Proof of Lemma \ref{lemma:upper bound}}\label{appendix:proof lemma ub}
Consider Problem \eqref{prob:full CSI} with $\Sete=\Seti=\mathcal{N}',k\in\mathcal{K}$. Since $\Sete=\mathcal{N}'$, from \eqref{eq:mk} and \eqref{eq:best SC}, $m(k)=\tilde{\Pi}(k),k\in\mathcal{K}$. By Proposition \ref{proposition:A}, we have $q_{k,n}^\ast=0,n\neq \tilde{\Pi}(k),k\in\mathcal{K}$ for Problem \eqref{prob:full CSI} with $\Sete=\Seti=\mathcal{N}',k\in\mathcal{K}$. It follows that Problem \eqref{prob:full CSI} with $\Sete=\Seti=\mathcal{N}',k\in\mathcal{K}$ achieves same rate as Problem \eqref{prob:full CSI re1} with $\Pi(k)=\tilde{\Pi}(k),\Seti=\mathcal{N}',k\in\mathcal{K}$. From Proposition \ref{proposition:Aa}, $q_k^\ast=0,k\in\tilde{D}^{\rm c}$ for Problem \eqref{prob:full CSI re1} with $\Pi(k)=\tilde{\Pi}(k),\Seti=\mathcal{N}',k\in\mathcal{K}$. It follows that Problem \eqref{prob:full CSI re1} with $\Pi(k)=\tilde{\Pi}(k),\Seti=\mathcal{N}',k\in\mathcal{K}$ achieves same rate as Problem \eqref{prob:full CSI re1} with $\Pi(k)$ given in \eqref{eq:prior WET SC} and $\Seti=\mathcal{N}',k\in\mathcal{K}$. Therefore, Problem \eqref{prob:full CSI} with $\Sete=\Seti=\mathcal{N}',k\in\mathcal{K}$ achieves same rate as Problem \eqref{prob:full CSI re1} with $\Pi(k)$ given in \eqref{eq:prior WET SC} and $\Seti=\mathcal{N}',k\in\mathcal{K}$.
This thus completes the proof of Lemma \ref{lemma:upper bound}.

\end{document}